\documentclass{PoS}

\usepackage{amssymb}	
\usepackage{lineno}
\usepackage{subfig}
\usepackage[square,numbers]{natbib}
\usepackage{rotating}

\title{MAGIC and MWL monitoring of the blazar TXS 0506+056 in the 2018/2019 season}

\ShortTitle{MAGIC TXS0506+056 2018/19}

\author{\speaker{K. Satalecka}$^{1}$, E. Bernardini$^{2}$, W. Bhattacharyya$^{1}$, M. Cerruti$^{3}$, V. Fallah Ramazani$^{4}$,
L. Foffano$^{2}$, S. Inoue$^{5}$, E. Prandini$^{2}$, Ch. Righi$^{6}$, N. Sahakyan$^{7}$,
F. Tavecchio$^{6}$ for the MAGIC Collaboration \footnote{\texttt{https://magic.mpp.mpg.de/}  For collaboration list see PoS(ICRC2019)1177
}\\
        $^{1}$DESY, 15738 Zeuthen, Germany; $^{2}$Universit\`a di Padova and INFN, I-35131 Padova, Italy; $^{3}$Universitat de Barcelona, ICCUB, IEEC-UB, E-08028 Barcelona, Spain; $^{4}$Finnish Centre of Astronomy with ESO (FINCA), University of Turku, FI-20014 Turku, Finland; $^{5}$ICRR, The University of Tokyo, 277-8582 Chiba, Japan; $^{6}$National Institute for Astrophysics (INAF), I-00136 Rome, Italy; $^{7}$ICRANet-Armenia at NAS RA, 0019 Yerevan, Armenia\\
        E-mail: \email{konstancja.satalecka@desy.de}}

\abstract{The gamma-ray blazar TXS 0506+056, was discovered in VHE gamma-rays
by the MAGIC telescopes in 2017 in a follow-up campaign of a high energy neutrino
event IceCube-170922A (IC+Fermi+MAGIC++, Science 361, eaat1378 (2018)). Subsequent
multivawelenght (MWL) observations
and theoretical modeling in a frame of hadro-leptonic emission confirmed that this
source could be a potential cosmic ray and neutrino emitter
(MAGIC Collaboration, Ansoldi et al., (2018)).
This is, by far, the most significant association between a high-energy neutrino
and an astrophysical source emitting gamma rays and X-rays. TXS 0506+056 is a key object
to help the astrophysics community to establish connections between high-energy neutrinos
and astrophysical sources. Accurate and contemporaneous MWL spectral measurements are essential ingredients to achieve this goal.
In the conference, we present the measurements
from the MAGIC and MWL monitoring of this source, spanning the time period
from November 2017 till February 2019. These include the lowest VHE gamma-ray emission state measured from this source so far as well as a flaring episode in December 2018. }

\FullConference{36th International Cosmic Ray Conference -ICRC2019-\\
		July 24th - August 1st, 2019\\
		Madison, WI, U.S.A.}

\begin{document}

\section{Introduction}
Active Galactic Nuclei (AGN), especially of the blazar class, dominate the very high energy (VHE; > 100 GeV) electromagnetic sky. Blazars, as all AGN are highly luminous sources, powered by a supermassive black hole. Additionally they display relativistic jets, one of which is pointed near the line of sight of the observer, which results in a relativistic boosting of the emission. Blazars are also considered as prime candidates for hadronic accelerators \cite{1993A&A...269...67M, Mannheim:1995mm, 1997ApJ...488..669H}.
In this scenario, their VHE photon emission can be explained as secondary radiation from interactions of accelerated protons and the surrounding photon or matter fields. Another by-product of those interactions are neutrinos. If observed, they should provide us with a smoking-gun signature of hadronic interactions.  

IceCube, a 1km$^{3}$ neutrino telescope located at the South Pole, has performed several extensive searches for neutrino emission from AGN~\cite{2015ApJ...807...46A}. So far, no strong correlation between the cataloged AGN and astrophysical neutrinos has been found, and blazar contributions to the all-sky astrophysical neutrino flux has been constrained at the level of 27\% \cite{2017ApJ...835...45A}. Nevertheless, the all-sky astrophysical neutrino flux shows an isotropic distribution, favouring an extragalactic origin and AGN still remain as promising neutrino sources.

TXS 0506+056, a bright gamma-ray emitting blazar, is of special interest for the hadronic accelerators case. On September 22, 2017 a 290-TeV neutrino event was detected by IceCube (IceCube-170922A) in spatial and temporal coincidence with an enhanced $\gamma$-ray emission state of this source~\cite{IceCube:2018dnn}. The significance of this coincidence was estimated to be at the 3$\sigma$ level. 
MAGIC was the first IACT to detect VHE gamma-rays from this object shortly after IceCube issued the alert. During the subsequent monitoring of the source the MAGIC collaboration and other alerted instruments collected a large multiwavelenght data set, essential for modeling and interpretation of the TXS 0506+056 emission mechanism. The MAGIC collaboration proposed an interpretation in the hadro-leptonic spine-layer emission framework \cite{2018ApJ...863L..10A}, where the results of the modeling strongly support the hypothesis of TXS 0506+056 being a neutrino and cosmic ray emitter.

The TXS 0506+056 case is, our most compelling evidence of hadronic emission in blazars to date. Therefore during the period from November 2017 till February 2019 MAGIC, together with MWL partners continued to monitor the source behaviour. In the conference, we present the data collected during this extended multivawelenght monitoring campaign,
including longer periods of low state emission, as well as a VHE $\gamma$-ray flaring episode in December 2018.


\section{MAGIC monitoring 2017/2019}

MAGIC observed the source TXS~0506+056 for a total of about 90~h between November 2017 and February 2019 within a dedicated monitoring program aimed at collecting a long-term data sample of the source. 
The analysis was performed on $\sim$80~h of good-quality data with zenith angle range between 22$^\circ$ and 50$^\circ$, 
using the MAGIC Analysis and Reconstruction Software \cite{2009arXiv0907.0943M, MAGICsens}. 

During most of the monitored period the source was in \textit{low state} ($\sim$75~h), with an average flux above 80 GeV 
about 10-15 times lower than the flux observed during the flare in October 2018.
This is the lowest VHE $\gamma$-ray emission level observed form this source so far. On December 1$^{st}$ and 3$^{rd}$ 2018 (MJD 58453 and 58455) an enhanced emission was observed with fluxes comparable with the flare detected by MAGIC in October 2017, shortly after the neutrino alert IceCube-170922A \cite{2018ApJ...863L..10A}. On December 4$^{th}$, the MAGIC collaboration issued an Astronomers Telegram (ATel \#12260) to encourage further MWL observations of TXS 0506+056. Several ToOs, including the X-ray and optical instruments were triggered. Table \ref{tab:flux} shows the summary of the TXS 0506+056 flux levels as measured by MAGIC. 
\begin{table}[h]
\small
\caption{MAGIC measurements of TXS 0506+056}

\label{tab_param}
\centering
\begin{tabular}{l|ccc}
\hline
 Data set & Duration [h]& Significance & VHE activity\\
 \hline
 MJD 58453 & 2.5 & 3.8$\sigma$ &  High \\
 MJD 58455 & 1.8 & 5.4$\sigma$ &  Very high \\
 Rest & 74.4 & 4.0$\sigma$ &  Low \\
 \end{tabular}\label{tab:flux}
\end{table}


\section{Multiwavelenght observations} 
The preliminary multi-band fluxes that were shown at the conference were collected from public archives of each instrument. Large part of those observations were coordinated with MAGIC in order to ensure simultaneous exposure. Dedicated ToO observations were performed with Swift/XRT \cite{2004SPIE.5165..201B} and UVOT as well as NuSTAR \cite{harrison13}. The MAGIC observations were usually accompanied by the KVA \cite{2018A&A...620A.185N} optical telescope, additional measurements were performed with REM after the flare in December 2018. TXS 0506+056 is also systematically monitored by the ASAS-SN project \cite{2017PASP..129j4502K}.

\begin{figure}
\centering
\includegraphics[width=0.5\columnwidth]{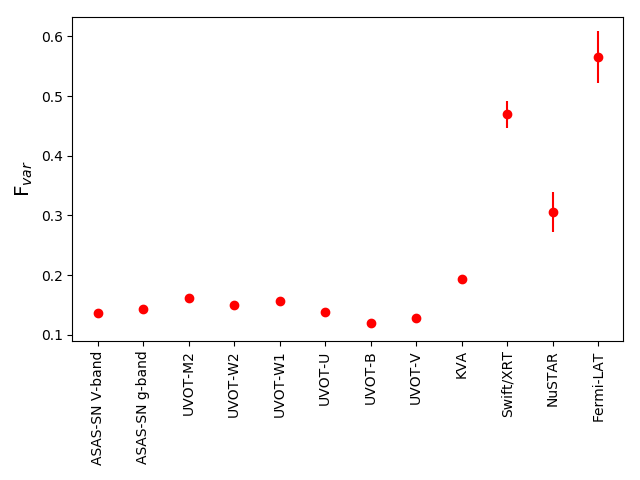}
\caption{Fractional variability parameter for each instrument}
\label{fig:mwl_lc}
\end{figure}

Clear variability is observed in all wavelengths. In order to quantify it, we calculated the {\it fractional variability parameter} F$_{var}$ according to \cite{2003MNRAS.345.1271V}. 
The most pronounced variability F$_{var}\sim$0.3-0.5 is observed in the highest energies (X-rays and $\gamma$-rays). The optical and UV bands display a moderate variability of F$_{var}\sim$0.1-0.2. As for VHE gamma rays, the very low activity during these two years yielded low significance flux measurements with MAGIC for most of the single night observations, which precluded to compute a meaningful fractional variability in this band.


The two distinct activity levels at VHE gamma rays motivated us to produce two broadband SEDs, for the low state and the flaring state, and we modeled this emission using a lepto-hadronic theoretical scenario as discussed in \cite{icrc:matteo}.
The largest contemporaneous MWL data sample was collected for the VHE $\gamma$-ray low emission state. Its spectral energy distribution shows a clear similarity in shape and flux level at the low energy peak to the one previously observed \cite{2018ApJ...863L..10A}. The flux level of the high energy peak was measured to be lower than before. 
For the flaring episode in December 2018, the VHE $\gamma$-ray flux is compatible with the previously measured flares. 


\section{Summary and Outlook}
TXS 0506+056 is a key object
to help the astrophysics community to establish connections between high-energy neutrinos
and astrophysical sources. Accurate and contemporaneous MWL spectral measurements are essential
ingredients to achieve this goal.
In the conference, we presented the results
from the MAGIC and MWL monitoring of this source, spanning the time period
from November 2017 till February 2019.

In comparison to the previously published results, TXS 0506+056 displayed a very low VHE $\gamma$-ray emission state during most of the observed nights, with the exception of the flaring activity observed on December 1$^{st}$ and 3$^{rd}$ 2018. 
The MWL light curve shows clear signs of variability, especially in the high energy range (X-ray and HE $\gamma$-ray), as quantified by the fractional variability parameter F$_{var}$.

We plan to perform a dedicated, low energy optimized analysis of the MAGIC data with a goal of recovering the signal below the currently obtained energy threshold. These results, along with an updated theoretical interpretation will be presented in an upcoming publication.

\section{Acknowledgments}
The authors gratefully acknowledge financial support from the agencies and organizations listed here: https://magic.mpp.mpg.de/acknowledgments\_ICRC2019/

This project has received funding from the European Union's Horizon2020 research and innovation programme under the Marie Sklodowska-Curie grant agreement no 664931.

\bibliographystyle{ICRC}
\bibliography{biblio}

\end{document}